\documentstyle[prl,aps,epsf,here]{revtex} 
\twocolumn 

\def\eq#1{(\ref{#1})}
\def\fig#1{Fig.~\ref{#1}}
\def\be{\begin{equation}}
\def\ee{\end{equation}}
\def\bea{\begin{eqnarray}}
\def\eea{\end{eqnarray}}
\def\bgrk#1{\mbox{{\boldmath $#1$ \unboldmath}}\!\!}
\def\siml{\,\hbox{\kern.1em \lower.6ex \hbox{$\sim$} \kern-1.12em
          \raise.6ex \hbox{$<$} }}
\def\simg{\,\hbox{\kern.1em \lower.6ex \hbox{$\sim$} \kern-1.12em
          \raise.6ex \hbox{$>$} }}
\newcommand{\Figurebb}[9]{
\begin{figure}[H]\begin{center}
\leavevmode
\epsfysize=#7cm
\epsfbox[#2 #3 #4 #5]{#6}
\par
\parbox{#8cm}{
\caption[figure]{\renewcommand{\baselinestretch}{0.8} \small
                                           \hspace{-0.3truecm}#9}
\label{#1}}\end{center}
\end{figure}
}

\begin{document}
\draft

\title{Semiclassical theory of spin-orbit interactions using spin coherent
       states\cite{byline}}
\author{M. Pletyukhov$^1$, Ch. Amann, M. Mehta$^2$ and M. Brack}
\address{Institut f\"ur Theoretische Physik, Universit\"at Regensburg,
         D-93040 Regensburg, Germany\\
         $^1$e-mail: {\rm mikhail.pletyukhov@physik.uni-regensburg.de}\\
         $^2$present address: Harish-Chandra Research Institute, 
         Chhatnag Road, Jhusi, Allahabad 211019, India}
\date{\today}
\maketitle

\vspace*{-0.62cm}

\begin{abstract}
We formulate a semiclassical theory for systems with spin-orbit
interactions. Using spin coherent states, we start from the
path integral in an extended phase space, formulate the classical
dynamics of the coupled orbital and spin degrees of freedom, and
calculate the ingredients of Gutzwiller's trace formula for the 
density of states. For a two-dimensional quantum dot with a spin-orbit 
interaction of Rashba type, we obtain satisfactory agreement with fully 
quantum-mechanical calculations. The mode-conversion problem, which 
arose in an earlier semiclassical approach, has hereby been overcome.
\end{abstract}


\vspace*{-0.25cm}

The incorporation of spin degrees of freedom in the semiclassical
treatment of quantum systems represents a considerable challenge to the
theorist. The periodic orbit theory (POT), initiated by Gutzwiller
through his semiclassical trace formula \cite{gutz}, has been very
successful in promoting the research on quantum chaos \cite{chao} and
in explaining prominent quantum shell effects occurring in finite
fermion systems in many domains of physics (see, e.g., \cite{book}).
The growing interest in ``spintronics'', i.e., the spin-polarized
transport and spin dynamics in various electronic materials
\cite{sptr}, with the particular scope of developing a spin transistor
\cite{tran}, make it desirable to formulate also the semiclassical
theory including spin.

Littlejohn and Flynn have extended the POT for systems with
multi-component wavefunctions and used it for the WKB quantization of
spherical systems with spin-orbit interactions \cite{lifl}. They did
not give an explicit trace formula, however, the main obstacle being
the problem of ``mode conversion'' which arises when the spin-orbit
interaction in phase space locally becomes zero. Frisk and Guhr
\cite{frgu} have applied the method of \cite{lifl} to deformed cavities
and promoted the hypothesis that spin-flips occur at the
mode-conversion points through diabatic transitions between the two
adiabatic spin-polarized energy surfaces. Bolte and Keppeler
\cite{boke} have derived a relativistic trace formula from the
Dirac equation and studied nonrelativistic limits, thereby justifying
some {\it ad hoc} assumptions made in \cite{frgu}. Their approach 
works well in the weak-coupling limit where the spin does not affect
the orbital motion \cite{wink}. It may not be extensible \cite{cham},
however, to strong spin-orbit interactions such as that observed in 
$p$-InAs or in InGaAs-InAlAs heterostructures \cite{inas}, or also that 
amongst the nucleons in atomic nuclei \cite{nucl}. For such systems,
the approach presented below provides a semiclassical theory with a 
larger range of validity.

In this letter, we report on a semiclassical approach that allows
for the explicit coupling between spin and orbital dynamics. We use the
spin coherent state method and the path integral in an extended phase
space to derive classical equations of motion which include the spin
degrees of freedom. We then obtain the amplitudes in the semiclassical
trace formula for the density of states without encountering the
mode-conversion problem.

The path integral for a system with spin in the SU(2) spin coherent
state representation originally appeared in a paper by Klauder
\cite{klau} as an integral on $S^2$. Kuratsuji {\it et al.}\
\cite{kura} have represented it as an integral over paths in the
extended complex plane $\bar{C}^1$. From the group theoretical point of
view, the SU(2) coherent state can be associated with the unitary
irreducible representation of the SU(2) group \cite{pere}. To present
its construction, we follow a recent paper by Kochetov \cite{koch} and
his notations. The coherent state $|z,S\rangle$ for spin $S$ is defined 
by
\bea
&~& |z; S \rangle = (1 + |z|^2)^{-S} \exp(z\hat{s}_{+})\, |S,-S\rangle\,,\\
&~& \hat{s}_{-}|S,-S\rangle = 0\, ,
\eea
where $z \in \bar{C}^1$ is a complex number, and
$\hat{s}_{\pm}=\hat{s}_1 \pm i\hat{s}_2$, and $\hat{s}_3$ are the
generators of the spin SU(2) algebra:
\be
[\hat{s}_3,\hat{s}_{\pm}] = \pm\hat{s}_{\pm}\,,\qquad
[\hat{s}_{+},\hat{s}_{-}] = 2\hat{s}_3\,.\qquad
\ee
$S \in N/2$ denotes the representation index.

The irreducibility as well as the existence of the group invariant de
Haar measure $d \mu_S$ ensures that the resolution of unity holds in
the spin coherent state basis:
\bea
\int \!|z;S\rangle\langle z;S| \,d\mu_S (z) = I_{2 S + 1}\,,\nonumber\\
d \mu_S (z) = \frac{2S+1}{\pi\,(1+|z|^2)^2} d^{\,2}z\,,
\eea
which turns out to be the most important property of spin coherent
states that allows for the path integral construction, whereby the
measure $d\mu_S$ takes account of the curvature of the sphere $S^2$. 
In what follows, we denote $|z;S\rangle$ simply by $|z\rangle$ and use
$z=u-iv$ with $u,v\in R^1$.

Let us now consider a quantum Hamiltonian with spin-orbit interaction
\be
\widehat{H} = \widehat{H}_0 (\hat{{\bf q}},\hat{{\bf p}})+\kappa\hbar\,
\hat{{\bf s}} \cdot \widehat{{\bf C}}({\hat{\bf q}},\hat{{\bf p}})\,,
\label{qh}
\ee
where $\hat{\bf s}=(\hat{s}_1,\hat{s}_2,\hat{s}_3)$. Hereby
$\widehat{\bf C}=(\widehat{C}_1,\widehat{C}_2,\widehat{C}_3)$ is a 
vector function of the coordinate and momentum operators $\hat{\bf q}$,
$\hat{\bf p}$, and the parameter $\kappa$ regulates the strength of the
interaction. We make use of Smirnov's idea \cite{smir} to write

\newpage

\noindent
the expression for the respective quantum propagator in terms of a path
integral in both the orbital variables ${\bf q,p}$ and the spin
coherent state variables $v,u$. Imposing periodic boundary conditions
on the propagation and thus integrating over closed paths, we arrive at 
the expression for the partition function (or trace of the propagator):
\be
Z (T) = \int \! {\cal D} [{\bf q}] {\cal D} [{\bf p}] {\cal D}
\mu_S [z] \,\exp\{i{\cal R}[{\bf q,p},z;T]/\hbar\},
\label{parti}
\ee
where ${\cal R}$ is Hamilton's principal action function, calculated
along closed paths over a time interval $T$ 
\bea
{\cal R} [{\bf q,p},z;T] & = & \!\oint_0^T \biggl[ \frac12
         ({\bf p}\!\cdot\!{\bf\dot q} - {\bf q}\!\cdot\!{\bf\dot p})
         + 2S\hbar\,\frac{(u{\dot v}-v{\dot u})}{(1\!+\!|z|^2)} \nonumber\\
     & & \hspace*{2.5cm} - \, {\cal H}({\bf q,p},v,u)\biggr] dt\,,
\label{princ}
\eea
and the path integration in \eq{parti} is taken over the
$2(d+1)$-dimensional extended phase space:
\[
{\cal D} [{\bf q}] {\cal D} [{\bf p}] {\cal D} \mu_S [z] = \lim_{N \to
\infty} \prod_{k=1}^{N} \prod_{j=1}^d \frac{d q_{j}(t_k) d p_{j}(t_k)}
                                        {2 \pi \hbar} d \mu_S (z_k)\,.
\]
Hereby the time interval $T$ is divided into $N$ time steps $t_k=kT/N$,
and $z_k=z(t_k)$. Note that the periodic boundary conditions enable the
antisymmetrization of the orbital part of the symplectic 1-form in
\eq{princ}.

The classical phase-space symbol ${\cal H}({\bf q,p},v,u)$ of the 
Hamiltonian (\ref{qh}), appearing in the integrand of \eq{princ}, is
\be
{\cal H}({\bf q,p},v,u)={\cal H}_0({\bf q,p}) +
\kappa\hbar S\, {\bf n}(v,u)\cdot {\bf C} ({\bf q}, {\bf p})\,,
\ee
where ${\cal H}_0({\bf q,p})$ and ${\bf C}({\bf q,p})$ are the Wigner
symbols of the operators $\widehat{H}_0$ and $\widehat{{\bf C}}$, and
${\bf n}=(n_1,n_2,n_3)=\langle z|\hat{\bf s}|z\rangle/S$ is the unit
vector of dimensionless classical spin components given in terms of $u$
and $v$ by $n_1=2u/(1+|z|^2)$, $n_2=2v/(1+|z|^2)$, and
$n_3=-(1-|z|^2)/(1+|z|^2)$.

The path integral \eq{parti} receives its largest contributions from
the neighborhood of the classical paths along which the principal
function ${\cal R}$ is stationary according to Hamilton's variational
principle $\delta {\cal R} = 0$. The first variation hereby yields the
following equations of motion
$$\dot{v} = \frac{(1\!+\!|z|^2)^2}{4\hbar S} \frac{\partial {\cal
H}}{\partial u}=\kappa \frac{(1\!+\!|z|^2)^2}{4} \, \frac{\partial {\bf n}
(v,u)}{\partial u} \cdot {\bf C} ({\bf q}, {\bf p})\,,\quad$$
$$\dot{u} = -\frac{(1\!+\!|z|^2)^2}{4\hbar S} \frac{\partial {\cal
H}}{\partial v}=- \kappa \frac{(1\!+\!|z|^2)^2}{4} \frac{\partial {\bf n}
(v,u)}{\partial v}\cdot{\bf C}({\bf q,p})\,,\!\!\!$$
$$\dot{q}_i = \frac{\partial {\cal H}}{\partial p_i} = \frac{\partial
{\cal H}_0}{\partial p_i} + \kappa\hbar S \, {\bf n} (v,u) \cdot
\frac{\partial {\bf C} ({\bf q}, {\bf p})}{\partial p_i} ,\qquad\qquad\quad$$
\be
\dot{p}_i = -\frac{\partial {\cal H}}{\partial q_i} = - \frac{\partial
{\cal H}_0}{\partial q_i} - \kappa\hbar S \, {\bf n} (v,u) \cdot
\frac{\partial {\bf C} ({\bf q}, {\bf p})}{\partial q_i},\qquad
\label{eom}
\ee
whose solutions define the classical orbits in the extended phase
space. Note the appearance of $\hbar$ multiplying $S$ everywhere in 
\eq{eom}, which reflects the non-classical nature of the spin. The 
two equations for the spin variables $v,u$ are equivalent to 
$\,\dot{{\bf n}}=-\kappa\,{\bf n} \times {\bf C}$, with the 
constraint ${\bf n}^2 = 1$. 
For spin-boson coupling in the Jaynes-Cummings model,

\newpage

\noindent
equations analogous to \eq{eom} have been derived in \cite{grab}. 

Sugita \cite{sugi} has recently given a re-derivation of Gutz\-willer's
trace formula, starting from $Z(T)$ whose Fourier-Laplace transform
yields the density of states:
\be
g(E) = \frac{1}{i\hbar}\int_0^\infty e^{iET/\hbar}\,Z(T)\,dT\,.
\label{gofE}
\ee
Following the general arguments of Gutzwiller \cite{gutz}, one may
evaluate the integrations in \eq{parti} using the stationary phase
approximation, which becomes exact in the classical limit ${\cal
R}\gg\hbar$. The semiclassical approximation of the partition function
$Z(T)$ then turns into a sum over all classical periodic orbits ($po$)
with fixed period $T$ 
\be
{\cal Z}_{sc}(T) = \sum_{po} e^{i{\cal R}_{po}/\hbar}\!
                    \int \!{\cal D}[\bgrk{\eta}]\,
                    \exp{\{i{\cal R}_{po}^{(2)}[\bgrk{\eta},T]/\hbar\}}\,,
\label{zscl}
\ee
where ${\cal R}_{po}$ are the principal functions \eq{princ}, evaluated 
now along the periodic orbits in the extended phase space. 
${\cal R}_{po}^{(2)}[\bgrk{\eta},T]$ are 
the second variations
\be
{\cal R}_{po}^{(2)}[\bgrk{\eta},T] = \oint_0^T \left[\frac12\bgrk{\eta} 
          \cdot {\cal J} \dot{\bgrk{\eta}}-{\cal H}^{(2)}\right]\!dt\,,
\ee
where $\bgrk{\eta}$ is the following $2(d+1)$-dimensional extended
phase-space vector of small variations
\[
\bgrk{\eta} = (\bgrk{\lambda},\nu,\bgrk{\rho},\xi)
            = \left(\delta{\bf q},
              \delta v\,\frac{2\sqrt{\hbar S}}{(1+|z|^2)},\delta{\bf p},
              \delta u\,\frac{2\sqrt{\hbar S}}{(1+|z|^2)}\right)\!,
\]
and {\small ${\cal J}=\left(\begin{array}{cc} 0 & I\! \\ \!\!-I & 0 
\end{array}\right)$}
is the $2(d+1)$-dimensional unit symplectic matrix. ${\cal H}^{(2)}$ is
the second variation of the classical Hamiltonian, calculated along
the periodic orbits:
\bea
{\cal H}^{(2)} & = & \frac{1}{2} \sum_{i,j} \left[\lambda_i \lambda_j
\frac{\partial^2{\cal H}}{\partial q_i \partial q_j} + 2 \lambda_i \rho_j
\frac{\partial^2{\cal H}}{\partial q_i \partial p_j} + \rho_i \rho_j
\frac{\partial^2{\cal H}}{\partial p_i \partial p_j}\right] \nonumber\\
   & + & \frac{\kappa}{2}(\nu^2+\xi^2)\,
         \left(-u\,C_1-v\,C_2+C_3\right) \nonumber\\
   & + & \kappa\sqrt{\hbar S}\,\frac{(1+|z|^2)}{2}
         \left(\nu\frac{\partial{\bf n}}{\partial v}
              +\xi\frac{\partial{\bf n}}{\partial u}\right) \nonumber\\
   & & \hspace*{2.6cm}
       \cdot \sum_{j}
       \left(\lambda_j \frac{\partial {\bf C}}{\partial q_j}
       +\rho_j\frac{\partial{\bf C}}{\partial p_j} \right)\!.
\eea
After a stationary-phase evaluation of the Fourier integ\-ral \eq{gofE}, 
one finally obtains Gutzwiller's trace formula in the standard form 
\cite{gutz,sugi}. What is new here is that all its ingredients are
obtained from the dynamics in the extended phase space: the actions 
${\cal S}_{po}(E)={\cal R}_{po}$+$ET_{po}$ (at fixed energy $E$) 
and the periods $T_{po}=dS_{po}/dE$ of the periodic orbits, the Maslov 
indices $\sigma_{po}$ for which Sugita has given general formulae 
\cite{sugi}, and the mono\-dromy matrix ${\cal M}_{po}$ defined by 
$\bgrk{\eta}(T_{po})={\cal M}_{po}\,\bgrk{\eta}(0)$ in terms of the 
solutions of the linearized equations of motion
$\dot{\bgrk{\eta}}={\cal J}\,\partial{\cal H}^{(2)}\!/\partial{\bgrk{\eta}}$.
We refer to a forthcoming extended paper \cite{plet} for the details of 
our calculations, where we also show that in the weak-coupling limit 
we obtain the same results as Bolte and Keppeler \cite{boke}.

\newpage

We shall presently illustrate our method by applying it to a simple
model Hamiltonian describing a two-dimensional electron gas $(S=1/2)$
in a semiconductor heterostructure, laterally confined to a quantum
dot, including a spin-orbit interaction of Rashba type \cite{rash}
\be
\widehat{H} = \frac{(\hat{p}_x^2 + \hat{p}_y^2)}{2m^*} + V(x,y)
 + \kappa\hbar\,(\sigma_y \hat{p}_x - \sigma_x \hat{p}_y)\,.
\label{hqdot}
\ee
Here $m^*$ is the effective mass of the electron, $\sigma_i$ are
the Pauli matrices, and the lateral confining potential $V(x,y)$ is 
approximated as a deformed harmonic oscillator
\be
V(x,y) = m^*(\omega_x^2\, x^2 + \omega_y^2\, y^2)/2\,.
\ee
We use $\omega_x=1.56\,\omega_0$, $\,\omega_y=1.23\,\omega_0$, and 
units such that $\hbar=m^*$ $=\omega_0=1$ and that $E$ and $\kappa$ become 
dimensionless. The spin-orbit coupling strength $\kappa$ in \eq{hqdot} 
depends on the band struc\-ture \cite{darn}. E.g., for a InGaAs-InAlAs 
quantum dot with $\sim$ 100 confined electrons one would obtain 
a value of $\kappa\sim 0.25$. We investigate in the following a 
situation where $\kappa$ is large enough for the spin dynamics to affect 
the orbital motion. For the examples below we have chosen $\kappa=0.67$. 
The equations of motion \eq{eom} were solved numerically.

For a large range of parameters with $0<\kappa\siml 0.75$, we find 
the following periodic solutions of Eq.\ \eq{eom}: 1) two pairs of
adiabatic orbits A$^\pm_x$ and A$^\pm_y$, librating along the $x$ 
and $y$ axes with fully polarized spin $n_y=\pm 1$ and $n_x=\pm 1$,
respectively; 2) two pairs of diabatic orbits, D$^\pm_{x1}$ and 
D$^\pm_{x2}$, oscillating around A$^\pm_x$ with $n_y\!\sim 0$; 
and 3) two pairs of diabatic orbits, D$^\pm_{y1}$ and D$^\pm_{y2}$, 
oscillating around A$^\pm_y$ with $n_x\!\sim 0$. The superscripts ($\pm$) 
of the diabatic orbits denote their senses of rotation. The stabilities
of these orbits will be discussed elsewhere \cite{plet}. For stronger 
coup\-lings $\kappa\simg 0.75$ and for energies $E\siml 8$, 
new orbits bifurcate from the A and D orbits. In these situations, the 
stability amplitudes will have to be regularized by suitable uniform 
approximations \cite{ssun}.

The diabatic orbits obtained in our approach reflect the explicit
coupling of the spin and orbital degrees of motion. Such orbits have
not been discussed in earlier semiclassical methods. The adiabatic
orbits with ``frozen spin'' correspond to those discussed in
\cite{lifl,frgu,cham} where they could not be used semiclassically,
though, because of the mode conversion occurring at their turning
points. We repeat that we do not encounter the mode-conversion
problem here, and that the stability amplitudes of all orbits can 
be readily calculated numerically.

In \fig{orbits} we present the $(x,y)$ shapes of the orbits 
A$^{\!+}_{x}$, D$^+_{x1}$, and D$^+_{x2}$ (left panels), and the 
time dependence of
their spin components $n_x$, $n_y$, and $n_z$ over one period (right 
panels), all evaluated at $E=60$. (The orbits A$^\pm_y$, D$^\pm_{y1}$, 
and D$^\pm_{y2}$ have analogous shapes concentrated along the $y$ 
axis, and the behavior of their $n_x$ and $n_y$ components is reversed.) 
We see that along the diabatic orbits D$^+_{x1}$ and D$^+_{x2}$, the 
spin rotates mainly near the $(n_x,n_z)$ plane (with $n_y\sim 0$)
in a non-uniform way. The diabatic 

\Figurebb{orbits}{3}{-5}{568}{705}{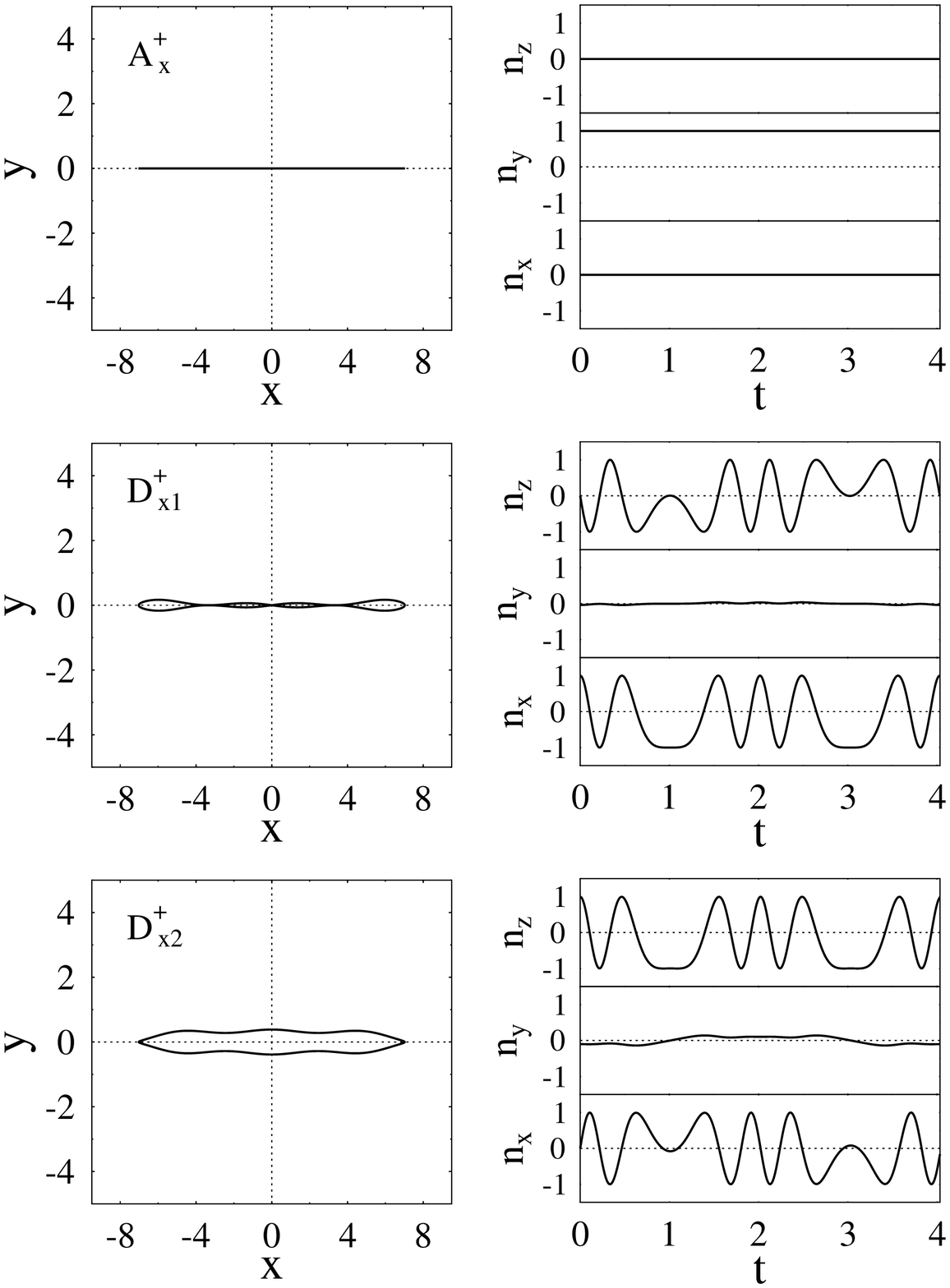}{11.3}{8.5}{
Periodic orbits in the 2-dimensional harmonic oscillator with Rashba
spin-orbit interaction (see text for param\-eters). {\it Left panels:}
orbits in the $(x,y)$ plane. {\it Right panels:} Spin components $n_x$,
$n_y$, and $n_z$ versus time. {\it From top to bottom:} Adiabatic orbit
A$^{\!+}_x$ along $x$ axis with polarized spin in $y$ direction, and
diabatic orbits D$^+_{x1}$ and D$^+_{x2}$ oscillating around the $x$
axis with spin rotating near the $(n_x,n_z)$ plane.
}

\noindent
spin-flip hypothesis of \cite{frgu} has thus been replaced here by 
the more sophisticated spin dynamics obtained from the 
coupled equations of motion \eq{eom}.

In \fig{levden} we show the oscillating part of the densi\-ty of states 
$\delta g(E)$, obtained quantum-mech\-anically (solid line) using exact 
diagonalization of the Hamilton\-ian \eq{hqdot}, and semiclassically 
(dashed line) using Gutzwiller's trace formula. Since the periodic 
orbit sum gener\-ally does not converge in systems with mixed classical 
dynamics, we have convoluted it with a normalized Gaussian, 
$\exp\{-(E/\gamma)^2\}/\gamma\sqrt{\pi}$. This brings the sum to
convergence \cite{sist}, whereby only the orbits with shortest periods
contribute, and reflects the prominent gross-shell structure 
of the quantum spectrum. We have used $\gamma=0.6$ where it was 
sufficient to include the above twelve primitive periodic orbits.
We observe a rather good agreement between the semiclassical and 
quantum-mechanical results. The regular beat-like structure comes
about through the interference of the six types of classical orbits which
all have frequencies close to either $\omega_x$ or $\omega_y$.

\newpage

\onecolumn

\Figurebb{levden}{30}{280}{795}{530}{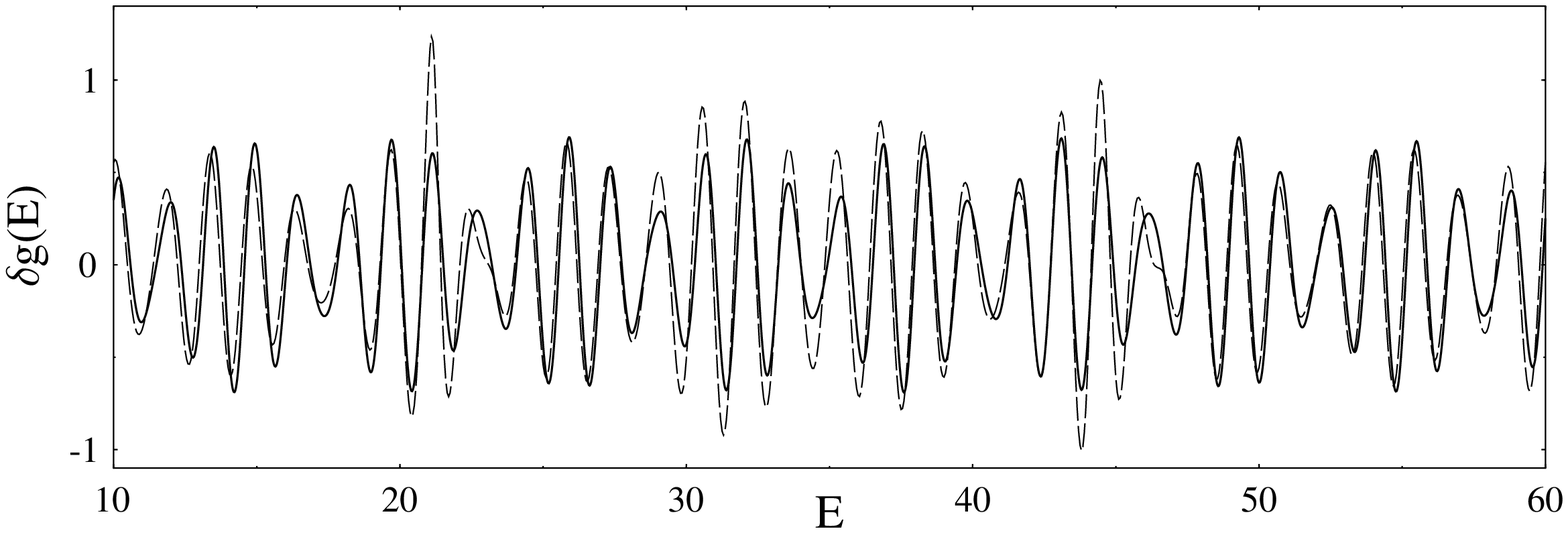}{5.2}{15}{
Oscillating part of the coarse-grained density of states (Gaussian
averaging parameter $\gamma=0.6$) versus energy for the same system as
in \fig{orbits}. {\it Solid line:} quantum-mechanical result, {\it
dashed line:} semiclassical result using the 12 primitive orbits.
}

\vspace*{-0.15cm}

In summary, we have presented a novel approach to include a non-adiabatic 
coupling of spin and orbital dynamics into the semiclassical theory of 
periodic orbits. Our approach overcomes some of
the difficulties with earlier approaches, in particular the restriction
to purely adiabatic spin motion and the problem of mode conversion. For
the simple model of a two-dimensional quantum dot with Rashba spin-orbit 
interaction, we obtain a satisfactory semiclassical description of the 
coarse-grained density of states. To study its experimental implications, 
we are in the process of calculating semiclassically the conductance of 
this system in response to an external magnetic field \cite{plet}.

We should point out that there exists a subtle problem connected with 
the fact that the measure ${\cal D}[\bgrk{\eta}]$ of the path integral
over the extended phase-space variations $\bgrk{\eta}$ in \eq{zscl} 
has the proper normalization only in the large-spin limit. For small 
spin, like $S=1/2$, this calls for an appropriate renormalization 
scheme \cite{miet}. Although the different renormalizations needed for 
pure spatial motion \cite{sola} and pure spin motion \cite{klau} are 
known, a valid scheme for the combined spin-orbit dynamics under 
investigation here is yet to be found. It might lead to extra phase 
corrections in the semiclassical expressions, as discussed in 
\cite{ston}, and thereby affect the numerical results to some extent. 

We acknowledge helpful and critical discussions with O. Zaitsev and
encouraging comments by K. Richter.

\end{document}